\documentclass[10pt,aps,prx,twocolumn,superscriptaddress]{revtex4-2}
\usepackage{amsmath,amssymb}
\usepackage{bm,mathtools}
\usepackage[breaklinks=true,colorlinks,citecolor=blue,linkcolor=blue,urlcolor=blue]{hyperref}
\usepackage[pdftex]{graphicx}
\usepackage{float}
\restylefloat{algorithm}
\usepackage{txfonts}
\usepackage{braket}
\usepackage{algorithm}
\usepackage{algpseudocode}
\usepackage{siunitx}
\usepackage{orcidlink}

\algdef{SE}[SUBALG]{Indent}{EndIndent}{}{\algorithmicend\ }%
\algtext*{Indent}
\algtext*{EndIndent}

\newcommand{\alglabel}[1]{Algorithm~\ref{#1}}

\newcommand{\figlabel}[1]{Fig.~\ref{#1}}
\newcommand{\seclabel}[1]{Sec.~\ref{#1}}

\DeclareMathOperator{\diag}{diag}
\newcommand{\trace}[1]{\text{tr}#1}

\makeatletter
\def\bibsection{%
   \par
   \begingroup
    \baselineskip26\p@
    \bib@device{\hsize}{72\p@}%
   \endgroup
   \nobreak\@nobreaktrue
   \addvspace{19\p@}%
  }%
\makeatother

\begin{document}
\title{Branch-and-bound digitized counterdiabatic quantum optimization} 

\author{Anton Simen$^{\orcidlink{0000-0001-8863-4806}}$}
\affiliation{Kipu Quantum GmbH, Greifswalderstrasse 212, 10405 Berlin, Germany}
\affiliation{Department of Physical Chemistry, University of the Basque Country UPV/EHU, Apartado 644, 48080 Bilbao, Spain}

\author{Sebastián V. Romero$^{\orcidlink{0000-0002-4675-4452}}$}
\affiliation{Kipu Quantum GmbH, Greifswalderstrasse 212, 10405 Berlin, Germany}
\affiliation{Department of Physical Chemistry, University of the Basque Country UPV/EHU, Apartado 644, 48080 Bilbao, Spain}

\author{Alejandro Gomez Cadavid$^{\orcidlink{0000-0003-3271-4684}}$}
\affiliation{Kipu Quantum GmbH, Greifswalderstrasse 212, 10405 Berlin, Germany}
\affiliation{Department of Physical Chemistry, University of the Basque Country UPV/EHU, Apartado 644, 48080 Bilbao, Spain}

\author{Enrique Solano$^{\orcidlink{0000-0002-8602-1181}}$}
\email{enr.solano@gmail.com}
\affiliation{Kipu Quantum GmbH, Greifswalderstrasse 212, 10405 Berlin, Germany}

\author{Narendra N. Hegade$^{\orcidlink{0000-0002-9673-2833}}$}
\email{narendrahegade5@gmail.com}
\affiliation{Kipu Quantum GmbH, Greifswalderstrasse 212, 10405 Berlin, Germany}

\date{\today}
\begin{abstract}
Branch-and-bound algorithms effectively solve combinatorial optimization problems, relying on the relaxation of the objective function to obtain tight lower bounds. While this is straightforward for convex objective functions, higher-order formulations pose challenges due to their inherent non-convexity. In this work, we propose branch-and-bound digitized counterdiabatic quantum optimization (BB-DCQO), a quantum algorithm that addresses the relaxation difficulties in higher-order unconstrained binary optimization (HUBO) problems. By employing bias fields as approximate solutions to the relaxed problem, we iteratively enhance the quality of the results compared to the bare bias-field digitized counterdiabatic quantum optimization (BF-DCQO) algorithm. We refer to this enhanced method as BBB-DCQO. In order to benchmark it against simulated annealing (SA), we apply it on sparse HUBO instances with up to $156$ qubits using tensor network simulations. To explore regimes that are less tractable for classical simulations, we experimentally apply BBB-DCQO to denser problems using up to 100 qubits on IBM quantum hardware. We compare our results with SA and a greedy-tuned quantum annealing baseline. In both simulations and experiments, BBB-DCQO consistently achieved higher-quality solutions with significantly reduced computational overhead, showcasing the effectiveness of integrating counterdiabatic quantum methods into branch-and-bound to address hard non-convex optimization tasks.
\end{abstract}

\maketitle

\section{Introduction}

Binary optimization problems are inherently challenging for modern computation, and this complexity increases when dealing with higher-order terms in the objective function or constraints~\cite{nonlinearprog}. The non-convexity in such problems complicates their solution, as traditional convex relaxations become ineffective~\cite{surveynonconvex, xu2024relaxationschallenges, danilova2022nonconvexchallenges, arjevani2020nonconvexchallenges, bertsekas2003convex}. Furthermore, the standard relaxation of integrality constraints no longer provides viable approximations, making these problems particularly difficult to solve efficiently \cite{surveynonconvex}, issues that are present in classical commercial solvers such as CPLEX~\cite{cplex} or Gurobi~\cite{gurobi}, among others. These binary optimization problems, whether constrained or unconstrained, can be readily mapped to two-local Ising spin-glass systems \cite{lucas2014ising}, which, despite their complexity, benefit from heuristics specifically designed to tackle such problems \cite{sim_annealing, tabu}. An increasing complexity can also be found in problems that are mapped to higher-order spin hamiltonians ~\cite{crisanti1992spherical, pelofske2023quantum, pelofske2024short-depth}, such as protein folding \cite{robert2021resource} and satisfiability problems \cite{boulebnane2022solving}. 
Recent developments have demonstrated progress in leveraging current quantum computers to solve combinatorial optimization problems \cite{abbas2024challenges, koch2025quantum, kotil2025quantum, boulebnane2022solving}.
Quantum annealing and its digital counterparts, such as the quantum approximate optimization algorithm (QAOA) \cite{farhi2014quantum}, are widely explored approaches for addressing these problems. However, these heuristic methods face significant challenges due to noise and the limited coherence times of contemporary quantum hardware. On the other hand, QAOA and its variants, despite benefiting from the flexibility of digital quantum computers, are constrained by inherent issues in variational quantum algorithms, such as barren plateaus and high resource demands~\cite{cerezo2024doesprovableabsencebarren,larocca2024reviewbarrenplateausvariational}.

To address these challenges, digital and purely quantum algorithms have been developed with the introduction of counterdiabatic (CD) Hamiltonians, which rely on approximations of the adiabatic gauge potential (AGP)~\cite{kolodrubetz2017geometry, sels2017minimizing, claeys2019floquet, hatomura2021controlling, takahashi2024shortcuts}. The purpose of CD Hamiltonians is to accelerate the annealing process by suppressing excitations between eigenstates, thereby improving the efficiency of the algorithm \cite{del2013shortcuts, hegade2021shortcuts, claeys2019floquet,demirplak2003adiabatic, berry2009transitionless}. Significant progress has recently been made in the field of digitized counterdiabatic quantum algorithms on tackling large-scale optimization problems \cite{hegade2022digitized, cadavid2024bias, romero2024bias}. These advancements have demonstrated robustness and scalability as purely quantum algorithms to tackle combinatorial optimization problems, while removing any dependence on classical optimizers. Notably, the bias-field digitized counterdiabatic quantum optimization (BF-DCQO) algorithm \cite{romero2024bias, cadavid2024bias, iskay} leverages ideas from quantum annealing with bias fields \cite{grass2019quantum,grass2022quantum} and iteratively refines the initial Hamiltonian based on observable measurements. This approach has shown that the family of solutions, represented in the resulting probability distribution, improves significantly with each iteration until reaching a saturation point, where the distribution includes multiple high-quality solutions. A substantial outperformance in solution quality compared to its analog counterpart was observed, specifically for higher-order binary optimization \cite{romero2024bias}. 

In this work, we propose a novel quantum branch-and-bound (BB) workflow for solving general higher-order unconstrained binary optimization (HUBO) problems, without relying on convex relaxations. This approach incorporates a BB strategy directly into the BF update rule and leverages quantum algorithms—such as BF-DCQO—as subroutines to generate approximate relaxed solutions. In particular, we utilize bias fields derived from z-basis projections of DCQO solutions to guide a structured and efficient exploration of the solution space. While we employ BF-DCQO in this work due to its demonstrated effectiveness on current quantum hardware, the workflow is modular and compatible with other quantum solvers. Specifically, BBB-DCQO constructs a binary search tree, where branches are generated based on biases with expectation values closest to zero—indicating high uncertainty (see~\figlabel{fig:schematics}). These expectation values, computed as longitudinal field measurements via $\langle \sigma^z \rangle$ observables, follow the approach proposed in \cite{cadavid2024bias}. Spins with high uncertainty are fixed in both possible orientations as longitudinal fields to create new branches, while a pruning strategy discards less promising paths to reduce computational costs. By introducing these spin constraints through bias fields, we do not remove spins from the target Hamiltonian but instead guide the optimization process while preserving quantum fluctuations. This is because the transverse field in the initial Hamiltonian remains active, allowing the system to explore different configurations and escape local minima.

We benchmarked BBB-DCQO against SA~\cite{kirkpatrick1983optimization} in two different regimes: (i) sparse 156-qubit HUBOs, which were simulated using tensor network-based methods, and (ii) dense 100-variable HUBOs IBM quantum hardware~\cite{ibm}, which become more challenging for tensor network simulation based on matrix product states (MPS).

For a more thorough comparison, we have also conducted both experiments using D-Wave quantum annealers~\cite{dwave}, known for solving QUBO problems naturally. In the quantum hardware experiments, we additionally employed a greedy local search algorithm to correct flip errors and adjust the energy distribution, improving the quality of the obtained solutions. In both simulated and hardware-based benchmarks, BBB-DCQO demonstrated superior performance, significantly reducing the number of function evaluations required to reach high-quality approximate solutions compared to SA. The SA algorithm used in this work is well suited for HUBO problems and does not require auxiliary variables.

The article is structured as follows: in~\seclabel{sec:formulation} we introduce the BBB-DCQO algorithm and the interplay between the parameters chosen and the expected performance. In~\seclabel{sec:results}, we present the HUBO instances and the different classical and quantum solvers tested, explaining how we conducted experiments on quantum hardware. Finally, the conclusions drawn from the presented results are given in~\seclabel{sec:conclusion}. Further notes and extended results are provided in the Appendices.
\begin{figure}[!tb]
    \centering
    \includegraphics[width=1\linewidth]{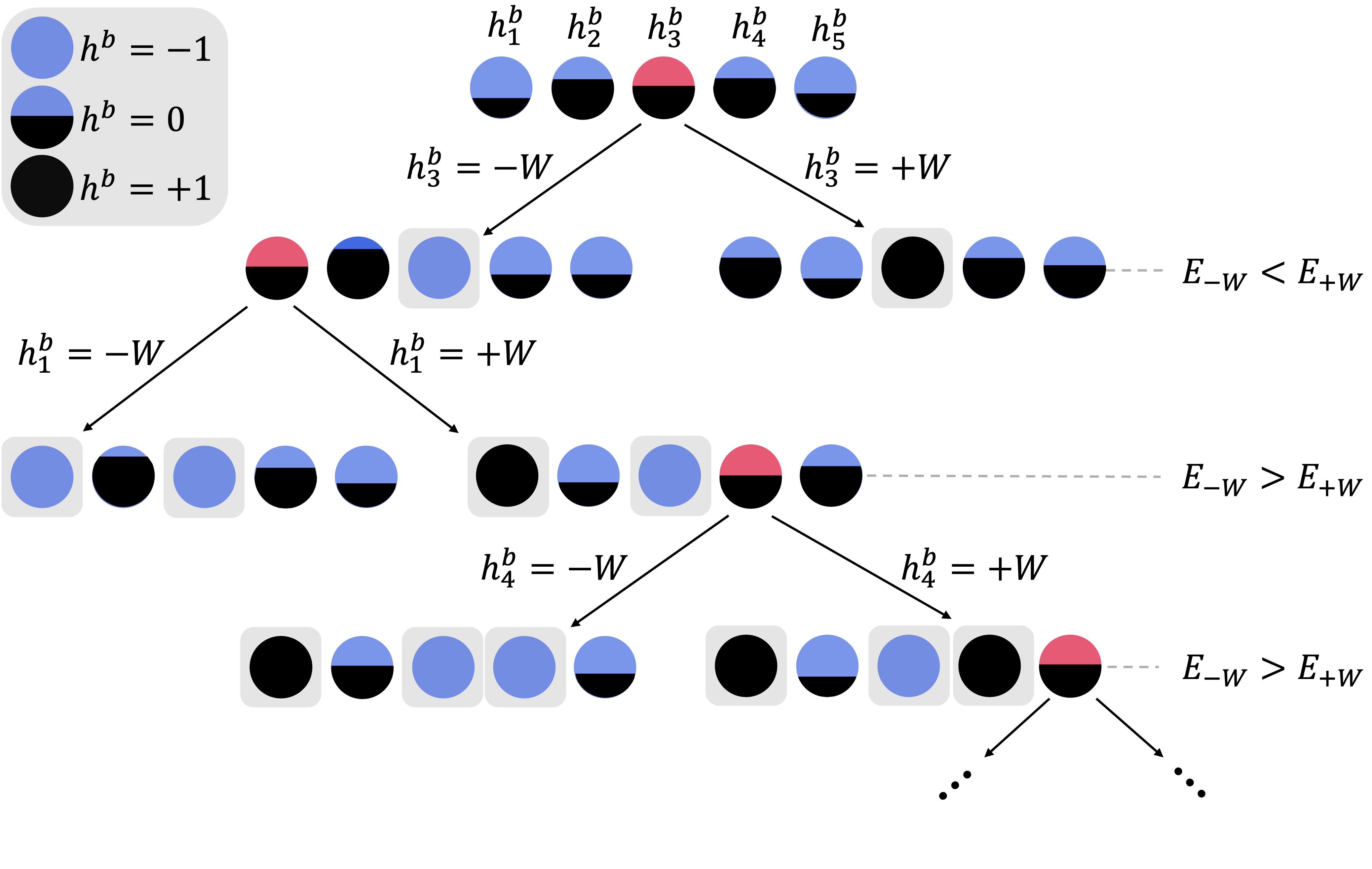}
    \caption{Schematics of the approximate BBB-DCQO algorithm. At the root, a set of bias fields $(h^b_1, \cdots, h^b_N)$ is obtained. Each bias field is represented as a partially filled blue circle. When the magnitude of a bias field is close to zero, the corresponding circle is half-filled and shown in red. At this point, two branches with magnitude $W$ and opposite signs are generated. The algorithm then selects the branch with the lowest energy for the next branching step.}
    \label{fig:schematics}
\end{figure}%

\section{DCQO with longitudinal bias fields}\label{sec:formulation}

Relying on the adiabatic theorem, adiabatic quantum optimization aims to evolve a system whose initial ground state is easy to prepare towards the ground state of a given problem Hamiltonian $H_f$. The Hamiltonian that will describe this process is $H_{\text{ad}}(\lambda)=  (1-\lambda) H_i + \lambda H_f$, with $\lambda(t)$ a scheduling function that toggles from zero to one in the time window $t\in[0,T]$. In the adiabatic limit, guaranteed by $\dot{\lambda}(t)\to 0$, this protocol reaches the ground state of $H_f$. Here, we consider 
\begin{equation}
    H_f=\sum_i h_i^z \sigma^z_{i} + \sum_{i<j} J_{ij} \sigma^z_{i} \sigma^z_{j}+ \sum_{i<j<k} K_{ijk} \sigma^z_{i} \sigma^z_{j} \sigma^z_{k}
\end{equation}
a $p$-spin glass with $p=3$~\cite{crisanti1992spherical}, which is also commonly known in quantum optimization as a higher-order Ising chain~\cite{pelofske2023quantum,pelofske2024short-depth}, encoding our HUBO problem. Hereinafter, we use $\lambda(t)=\sin^2\left(\frac{\pi}{2}\sin^2\frac{\pi t}{2T}\right)$ and $H_i=\sum_j(h^x_j\sigma^x_j + h^b_j\sigma^z_j)$, whose ground state is known for arbitrary $h^x_j$ ($h^b_j$), transverse- (longitudinal-)field contributions acting on the $j$-th spin, and natural units ($\hbar\equiv 1$).

It is possible perform a fast finite-time evolution by introducing an auxiliary counterdiabatic driving contributions, which suppresses diabatic transitions~\cite{demirplak2003adiabatic, berry2009transitionless}, which takes the form $H_\text{cd}(\lambda)=H_\text{ad}(\lambda)+\dot{\lambda}A_\lambda$, with $A_\lambda$ the adiabatic gauge potential~\cite{kolodrubetz2017geometry}. Nevertheless, its implementation is impractical due to its many-body structure and the need to know the entire spectrum. Accordingly, approximate implementations have been proposed~\cite{kolodrubetz2017geometry,sels2017minimizing,claeys2019floquet,hatomura2021controlling,takahashi2024shortcuts}, where it is possible to expand the adiabatic gauge potential as a nested-commutator series up to order $l$, namely $A^{(l)}_\lambda=i\sum_{k=1}^l \alpha_k(t) \mathcal{O}_{2k-1} (t)$, with $\mathcal{O}_{0}(t) = \partial_\lambda H_{\text{ad}}$ and $\mathcal{O}_{k}(t) = [ H_{\text{ad}}, \mathcal{O}_{k-1}(t) ]$. In the limit $l\to\infty$, it converges to the exact gauge potential. The coefficients $\alpha_k$ are obtained by minimizing the action $S_l=\trace{[G_l^2]}$ with $G_l=\partial_\lambda H_\text{ad} - i\big[H_\text{ad},A^{(l)}_\lambda\big]$. Hereinafter, we set $l=1$.

In the fast evolution regime, the adiabatic term $H_\text{ad}$ can be omitted, reducing the quantum resources needed without compromising performance~\cite{romero2024optimizing, chandarana2023digitized, dalal2024digitizedcounterdiabaticquantumalgorithms, cadavid2024bias, romero2024bias}. This approach is used in our studies, where only the CD contribution is implemented. To evolve on time such systems on analog quantum platforms remains a challenge mainly due to nonstoquasticity~\cite{hormozi2017nonstoquastic}. To overcome this issue, digitized counterdiabatic quantum protocols have been proposed for digital quantum computers~\cite{hegade2021shortcuts}. The resulting time-evolution operator can be decomposed into gates, up to $n_\text{trot}$ Trotter steps, as $U(T)=\prod_{k=1}^{n_{\text{trot}}} \prod_{j=1}^{n_\text{terms}} \exp [-i \Delta t \gamma_j(k \Delta t) H_j]$, with $H_\text{cd}=\sum^{n_\text{terms}}_{j=1}\gamma_j(t)H_j$ decomposed in $n_\text{terms}$ different $H_j$ operators with $\Delta t=T/n_\text{trot}$. Here, we use one effective Trotter step to preserve fidelity and not go beyond the coherence time of the quantum hardware.

Initialization plays a crucial role in the success of adiabatic quantum optimization and DCQO. Building upon the DCQO algorithm, instead of starting with an arbitrary initial state we can use BF-DCQO~\cite{cadavid2024bias,romero2024bias,iskay}, where the solution from DCQO is fed back as a bias to the input state for the next iteration. The total Hamiltonian, which includes the longitudinal bias fields, is defined as $H(\lambda)=  (1-\lambda) \tilde {H_i} + \lambda H_f + \dot{\lambda} A_{\lambda}^{(l)}$ with $\tilde{H}_i = \sum_{i=1}^{N} (h_i^x \sigma^x_i + h_i^b \sigma^z_i )$, where the value of the longitudinal bias field, \( h_i^b = \langle \sigma^z_i \rangle \), is a function based on the bias updating chosen, taking as input the Pauli-Z expectation values obtained by measuring $n_\text{shots}$ times the qubits in the computational basis after each iteration, where each sample is defined as $E_k$. We take $E= (1/n_\text{shots})\sum_{k=1}^{n_\text{shots}} E_k$ as the sampled mean energy, which estimates the expectation value of $H_f$. However, rather than using the entire sampled distribution ($\alpha=1$), it has been shown that using its lowest-energy-valued part is beneficial for faster convergence~\cite{Barkoutsos2020improving,barron2023provableboundsnoisefreeexpectation,romero2024bias}. Accordingly, we first sort them in terms of energy, such that $E_k\le E_{k+1}$ and then update the biases using a fraction $0<\alpha<1$ of the lowest-energy outcomes taken from $E(\alpha)= (1/\lceil \alpha n_\text{shots}\rceil)\sum_{k=1}^{\lceil \alpha n_\text{shots}\rceil} E_k$. 

Since the solution from the first iteration of DCQO is expected to sample low-energy states of the spin-glass Hamiltonian, $\langle \sigma^z \rangle$ can serve as an effective bias for the next iteration, steering the dynamics toward the actual solution. Because the bias term alters the initial Hamiltonian and, consequently, the initial ground state, the new ground state \( |\tilde{\psi_i}\rangle \) must be used as the input for the next iteration. The smallest eigenvalue of the single-body operator \( \big[h_i^x \sigma^x_i + h_i^b \sigma^z_i \big]  \) is given by \( \lambda^{\min}_i = -\sqrt{(h^b_i)^2 + (h^x_i)^2} \), with associated eigenvector \( \ket{\tilde{\phi}}_i = R_y(\theta_i) \ket{0}_i \). Since \( h^b_i \) and \( h^x_i \) are projections on the z- and x-axes, the corresponding ground state can be prepared by a y-axis rotation given by \( \theta_i = 2\tan^{-1}\left(\frac{-h^b_i + \lambda^{min}_i}{h^x_i}\right) \). Therefore, the ground state of \( \tilde{H_i} \) can be prepared using $N$ y-axis rotations as \( |\tilde{\psi_i}\rangle = \bigotimes_{i=1}^{N} |\tilde{\phi} \rangle_i = \bigotimes_{i=1}^{N} R_y(\theta_i)|0\rangle_i \). 

\subsection{Branch-and-bound BF-DCQO}

\begin{figure}[!t]%
\vspace{-2.7mm}
\begin{minipage}[t]{\columnwidth}%
\begin{algorithm}[H]%
    \begin{algorithmic}[1]
        \State \textbf{Input:} HUBO problem $P$ with objective function $F(z)$.
        \State \textbf{Initialize:} Set the best known solution $z^*$, and upper bound $UB = F(z^*)$.
        \State \textbf{Relax HUBO:} Get a continuous solution, $\tilde{z}$, from the relaxation.
        \State \textbf{Solve BF-DCQO:} Warm start BF-DCQO with $\bm{h^b} \leftarrow\tilde{z}$ and get a feasible solution $z$.
        \State Create a branching tree (BT) and insert the root node with $z$ and bound $LB = F(\tilde{z})$.

        \If{$F(z) < UB$}
            \State $z^* \leftarrow z$, $UB \leftarrow F(z)$
        \EndIf
        
        \While{BT has unexplored nodes}
            \State Select the node with the lowest lower bound $LB$ from BT.
            \State Get a relaxed solution, $\tilde{z}$ from the current subproblem.
            \State Assign $LB = F(\tilde{z})$
            \State Run BF-DCQO on the relaxed problem to obtain a spin solution $z$. 
            
            \If{$LB \ge UB$}
                \State Prune the branch and continue.
            \ElsIf{$z$ is a feasible spin solution and $F(z) < UB$}
                \State Update $z^* \leftarrow z$ and $UB \leftarrow F(z)$
            \ElsIf{$z$ is infeasible}
                \State Prune the branch and continue.
            \Else
                \State Get $i = \text{argmin}\left(|\tilde{z}_i|\right)$
                \State Create two subproblems with $z_i = \pm1$
                \State For each subproblem, relax and compute $LB = F(\tilde{z})$
                \State Insert each subproblem into BT with its $LB$
            \EndIf
        \EndWhile
        \State \textbf{Output:} Exact spin solution $z^*$.
    \end{algorithmic}
\caption{Exact BBB-DCQO}\label{alg:exact_bbb}%
\end{algorithm}%
\end{minipage}%
\end{figure}%
\begin{figure}[!t]%
\vspace{-7.55mm}
\begin{minipage}[t]{\columnwidth}
\begin{algorithm}[H]
    \begin{algorithmic}[1]
        \State \textbf{Input:} HUBO problem $P$ with objective function $F(z)$.
        \State \textbf{Initialize:} Set the best known solution $z^*$.
        \State \textbf{Run BF-DCQO:} Warm start BF-DCQO with $\bm{h^b_\text{init}} \leftarrow z^*$ and get: a feasible solution $z$; and new bias fields $\bm{h^b} = \left(\langle \sigma^z_0 \rangle, ... ,\langle \sigma^z_{N-1} \rangle\right)$.
        \State Create a branching tree (BT) and insert the root solution, $z$.

        \If{$F(z) < F(z^*)$}
            \State $z^* \leftarrow z$
        \EndIf
        
        \While{Number of layers $\leq K$}

            \State Select $i = \text{argmin}\left(|\bm{h^b} |\right)$
            \State Create two subproblems defined by $\bm{h^b_{\pm}} = \bm{h^b}$, with $\left(\bm{h^b_{\pm}}\right)_i = \pm W$

            \State \textbf{Run} BF-DCQO warm-started with both $\bm{h^b_-}$ and $\bm{h^b_+}$. Get the feasible solutions $z_{\pm}$ and their respective metrics $F_\pm=F(z_\pm)$.

            \State \textbf{Select} the node with best objective $F$, which gives $z_{\text{best}}$. Prune the other node.

            \If{$F(z_\text{best}) < F(z^*)$}
                \State $z^* \leftarrow z_\text{best}$
            \EndIf

        \EndWhile
        \State \textbf{Output:} Approximate solution $z^*$ found.
    \end{algorithmic}
\caption{Approximate BBB-DCQO}\label{alg:approx_bbb}
\end{algorithm}%
\end{minipage}%
\end{figure}%

In order to bias the solution of the binary optimization problem towards promising solutions and explore the search space more efficiently, we propose a branch-and-bound algorithm to update the bias fields. Branch-and-bound algorithms generally exploit convex relaxations to obtain continuous solutions, which will be then used as a resource to create subproblems based on additional constraints \cite{clausen1999branch}. We propose a version designed to tackle arbitrary problems, including those with non-convex higher-order cost functions. We consider the bias fields, $\bm{h^b} = \left(\langle \sigma^z_0 \rangle, ..., \langle \sigma^{z}_{n-1} \rangle\right)$, obtained from a (BF-)DCQO solution as a relaxation of the problem, whose outcomes are used to make decisions on which variables to branch on. Such decision can be made based on the magnitude of the bias fields. Assume that after DCQO with $h^x_i=-1$ and $h^b_i=0$, the $k$-th qubit has the highest uncertainty, i.e. its bias field $h^b_k$ is the closest to zero. Then, the initial Hamiltonian turns to
\begin{equation}
    \tilde{H_i} =\pm W\sigma^z_k + \sum_{i}h^x_i\sigma^x_i + \sum_{i \neq k}h^b_i\sigma^z_i,
\end{equation}
where $\pm W\sigma^z_k$ are two imposed constraints that break the uncertainty in the $k$-th qubit with magnitude $W$. Those two problems should be solved and evaluated based on a given metric. The magnitude of the bias, $W$, can be chosen based on prior knowledge about the problem, with $W=1$ the default value. Large values of $W$ make the longitudinal fields at the selected sites predominant over the transverse field, reducing the quantum fluctuations and, consequently, increasing the chances of getting stuck in local minima. This procedure is repeated following a binary tree structure, where the constraints on the bias fields remain on their corresponding branches. If there exists a reliable relaxation technique for HUBO problems, this method becomes exact, without guaranteed convergence in polynomial time. Conversely, the approximate version leverages a branch-and-cut approach, where non-promising branches are removed.

The exact version of BBB-DCQO follows from the standard branch‑and‑bound framework. It still needs an additional relaxation of the objective that can be solved quickly. \alglabel{alg:exact_bbb} gives the full procedure, assuming such a relaxation is available. The run stops when the usual optimality test of branch‑and‑bound is met, which certifies the global optimum. For users who care only about a guaranteed optimum, this variant offers no clear advantage over a fully classical solver, because the final check is still done classically. BBB-DCQO is most useful as a heuristic: BF-DCQO returns feasible binary solutions at every node, including the root, so high‑quality answers often emerge early in the search.

The approximate version of BBB-DCQO, described in Algorithm \ref{alg:approx_bbb}, lowers the computational cost by pruning any branch whose best energy is higher than that of its sibling, see Fig.~\ref{fig:schematics}. For a branching tree that extends over $K$ layers, BF-DCQO is executed $2K+1$ times. Because it limits the number of circuit executions, the approach fits current quantum devices, where running many quantum circuits is resource‑intensive. 

In this work, we focus on implementing the feasible \alglabel{alg:approx_bbb}, while leaving \alglabel{alg:exact_bbb} as a longer-term proposal, intended for future quantum computers and situations where the non-convex HUBO problems can be effectively relaxed.

\section{Results and discussion}\label{sec:results}

\begin{figure*}
    \centering
\includegraphics[width=\linewidth]{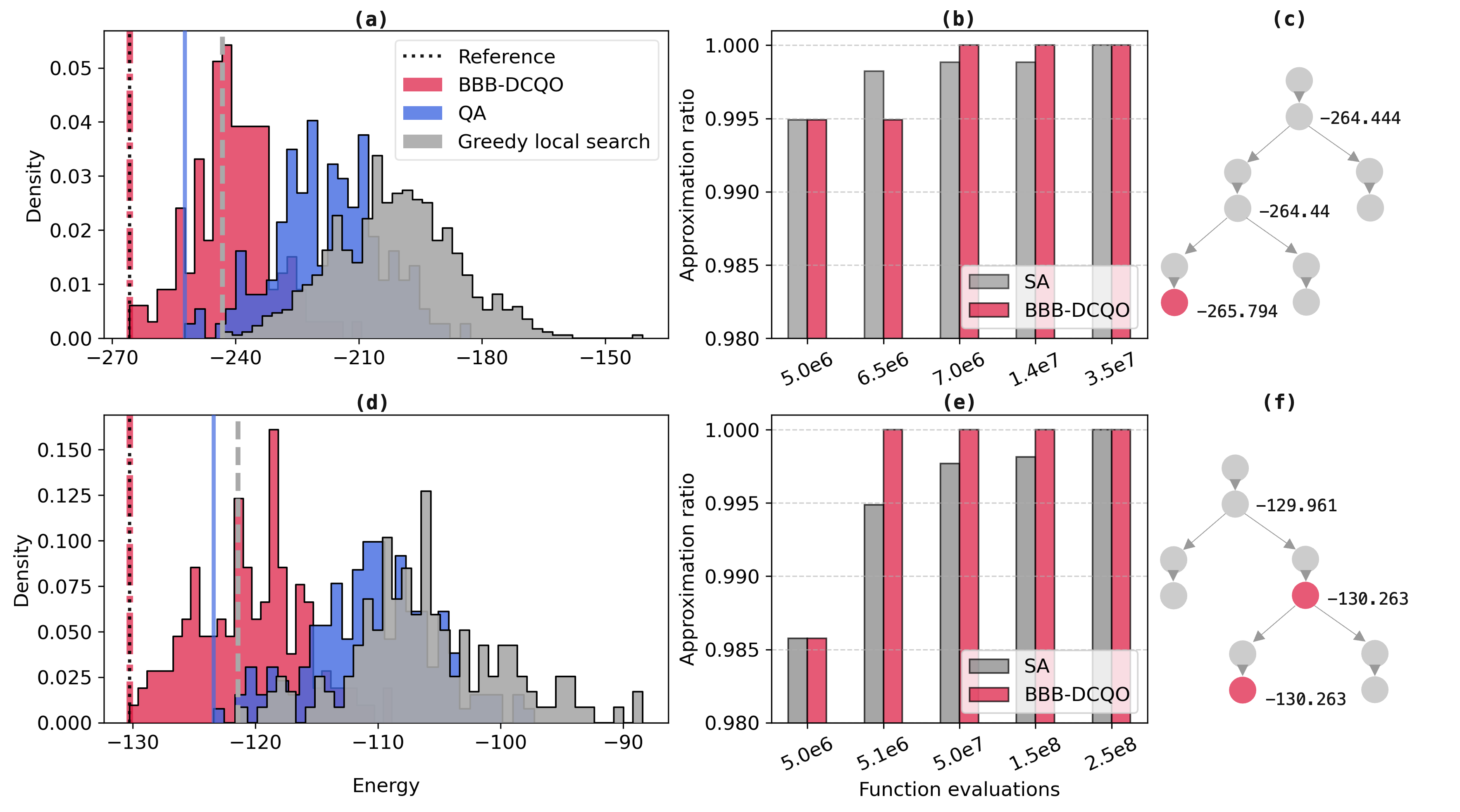}
    \caption{Experimental results on \textsc{ibm\_marrakesh} for two distinct 100-qubit HUBO instances. Figures (a) and (d) depict the final energy distributions obtained by BBB-DCQO, alongside the distributions resulting from QA with local search, and the pure greedy local search algorithm for both instances, constrained to the same number of function evaluations, namely $7.0\times 10^6$ for (a) and $5.1\times 10^6$ for (b). The reference solution \( e_{\text{ref}}\)  corresponds to the minimum energy achieved by SA with a large number of function evaluations, namely  \(5 \times 10^8\). The bar plots in panels (b) and (e) illustrate the approximation ratio, \( e_{\text{min}} / e_{\text{ref}} \), as a function of the number of function evaluations for both BBB-DCQO and SA. Here, function evaluations represent the number of energy measurements, which, for SA, corresponds to the number of spin-flips performed, while for BBB-DCQO, it accounts for both the number of quantum measurements and spin-flips applied in the warming-start SA and the greedy local search correction. Additionally, figures (c) and (f) illustrate the how BBB-DCQO converges to \(e_{\text{ref}} \) within the binary tree structure. The coloured nodes represent the moment where \( e_{\text{min}} / e_{\text{ref}} = 1\). 
}
    \label{fig:exp_res}
\end{figure*}

To assess the performance of BBB-DCQO under varying conditions, we structured the testing phase into two components. First, we executed hardware experiments on the \textsc{ibm\_marrakesh} quantum processor~\cite{ibm}, targeting two 100-qubit problem instances containing 182 and 241 two- and three-body terms, respectively. These dense instances contained couplings beyond nearest-neighbors on the hardware's heavy-hexagonal layout. As a result of the increasing density, simulating these circuits becomes more challenging for standard tensor network simulations based on MPS. The performance was assessed by comparing the minimum energy found by BBB-DCQO to that obtained via SA, with respect to the number of function evaluations required. BBB-DCQO was configured with a maximum tree depth of \(K=3\), using 2 iterations of BF-DCQO per node, including the root, and executing 10,000 shots per iteration. The bias magnitude of the branching variables was set to $W=30$, while the other bias fields were re-scaled to the range $|h^b_i| \leq 3$. A high local bias, $W$, in quantum hardware experiments is intended to mitigate noise from quantum fluctuations. Additionally, we introduced a greedy post-processing technique (see Appendix~\ref{appendix:post_processing}) for mitigating bit flip and measurement errors over the $150$ lowest-energy bitstrings with $15$ sweeps. This post-processing was applied after each circuit evaluation on the quantum hardware. The SA algorithm employed was natively compatible with higher-order unconstrained binary optimization (HUBO) problems, requiring no auxiliary variables or reformulations as in some available solvers, and utilized a geometric cooling schedule for temperature decay.

Secondly, we performed tensor network noiseless simulations on 156-qubit instances characterized by sparse connectivity, containing 155 and 154 two- and three-body terms, respectively. We compared the minimum energy obtained by BBB-DCQO against the best solutions from SA across two different sweep counts. The post-processing algorithm was not applied in these simulations to ensure a direct comparison of raw optimization performance.

\subsection{Hardware results}

Hardware results are presented in Fig.~\ref{fig:exp_res}. 
For both experiments, we warm-started BBB-DCQO using a solution obtained via SA with $500$ sweeps and $100$ reads, see Figs.~\ref{fig:exp_res}(b) and \ref{fig:exp_res}(e). This warm-start provides an initial bias that helps accelerate convergence in BBB-DCQO. Despite relying on SA for initialization, BBB-DCQO achieves lower-energy solutions with less resources for the studied instances. For the first instance, BBB-DCQO reaches the reference energy with $7.0 \times 10^6$ function evaluations at branching level $K = 2$ in the binary tree, against  $3.5 \times 10^7$ for SA. This total accounts for the function evaluations from SA as warm-start, quantum hardware shots, and evaluations performed during the greedy local search phase. For the second instance, BBB-DCQO identifies the reference solution at the first branching ($K = 1$) using $5.1 \times 10^6$ function evaluations, whereas SA requires approximately $2.5 \times 10^8$ evaluations to achieve the same result.

Figures~\ref{fig:exp_res}(a) and \ref{fig:exp_res}(d) depict the final solution distributions for BBB-DCQO, QA, and plain greedy local search over the uniform distribution. BBB-DCQO successfully recovers the reference solution in both cases, whereas greedy local search gets stuck in a local minimum. This observation highlights that greedy local search serves primarily as a local refinement heuristic and lacks the ability to discover high-quality solutions independently. Its effectiveness is amplified when used as an internal subroutine within BBB-DCQO. The QA solution quality is primarily affected by the  limitations of embedding higher-order terms on current quantum annealers. Specifically, for the QA results we use \textsc{Advantage\_system6.4}, a commercially available D-Wave quantum annealer~\cite{dwave}. To address HUBO instances there, a HUBO-to-QUBO mapping is required beforehand, introducing new variables containing products of the original ones as well as penalty terms to ensure that product constraints are met in the HUBO version of the problem. In particular, we set as penalty constants $15$ and $10$ for the instances solved in Figs.~\ref{fig:exp_res}(a) and~\ref{fig:exp_res}(b), respectively, which required $417$ and $320$ qubits in their QUBO encoding. These penalty constants were chosen because, heuristically, they yielded the best results during SA experiments after a HUBO-to-QUBO conversion, thus a better performance was expected on quantum hardware. Based on Ref.~\cite{pelofske2024short-depth}, we also set as annealing time $t_\text{a}=\SI{2000}{\micro\second}$, regime where the returned solutions are expected to be more optimal, and $n_\text{shots}=100000$, value that exceeds the resources used for our BBB-DCQO experiments. As a post-processing algorithm, we also applied \alglabel{alg:greedy} with the same parameters used in BBB-DCQO.

\subsection{Tensor network-based simulations}

The instances used for the noiseless simulations included one-dimensional nearest neighbor couplings. Thus, the circuits could be simulated by \texttt{matrix\_product\_state} simulator from Qiskit Aer~\cite{Qiskit}. SA was executed with $1000$ and $10000$ sweeps per run, each repeated over $100$ independent runs. Since each sweep involved one function evaluation per spin, this corresponds to a total of \(1.56 \times 10^7\) function evaluations for SA-$1000$ and \(1.56 \times 10^8\) for SA-$10000$. In contrast, BBB-DCQO was configured with a maximum tree depth of \(K=4\), using 3 iterations of BF-DCQO per node, including the root, and executing 15,000 quantum shots per iteration. The bias magnitude to the branching variables was set as $W=2$. This resulted in a total of \(4.05 \times 10^5\) function evaluations, which is approximately 38.5 times lower than SA-1000 and 385 times lower than SA-10000. The results shown in~\figlabel{fig:mps_30inst} demonstrate that BBB-DCQO achieves competitive or superior performance despite using significantly fewer function evaluations than both SA-1000 and SA-10000. Specifically, BBB-DCQO outperformed SA-1000 in 23 out of 30 instances (\(76.7\%\)), while SA-1000 achieved a better solution in 6 instances (\(20\%\)), with both methods obtaining the same energy in one instance. Against SA-10000, BBB-DCQO performed better in 7 instances (\(23.3\%\)), SA-10000 in 8 instances (\(26.7\%\)), and both methods reached the same energy in 15 instances (\(50\%\)). The structured search strategy of BBB-DCQO enables a more targeted exploration of the solution space, leveraging quantum sampling to efficiently navigate complex energy landscapes, whereas SA relies purely on stochastic updates.
\begin{figure}[!tb]
    \centering
    \includegraphics[width=\linewidth]{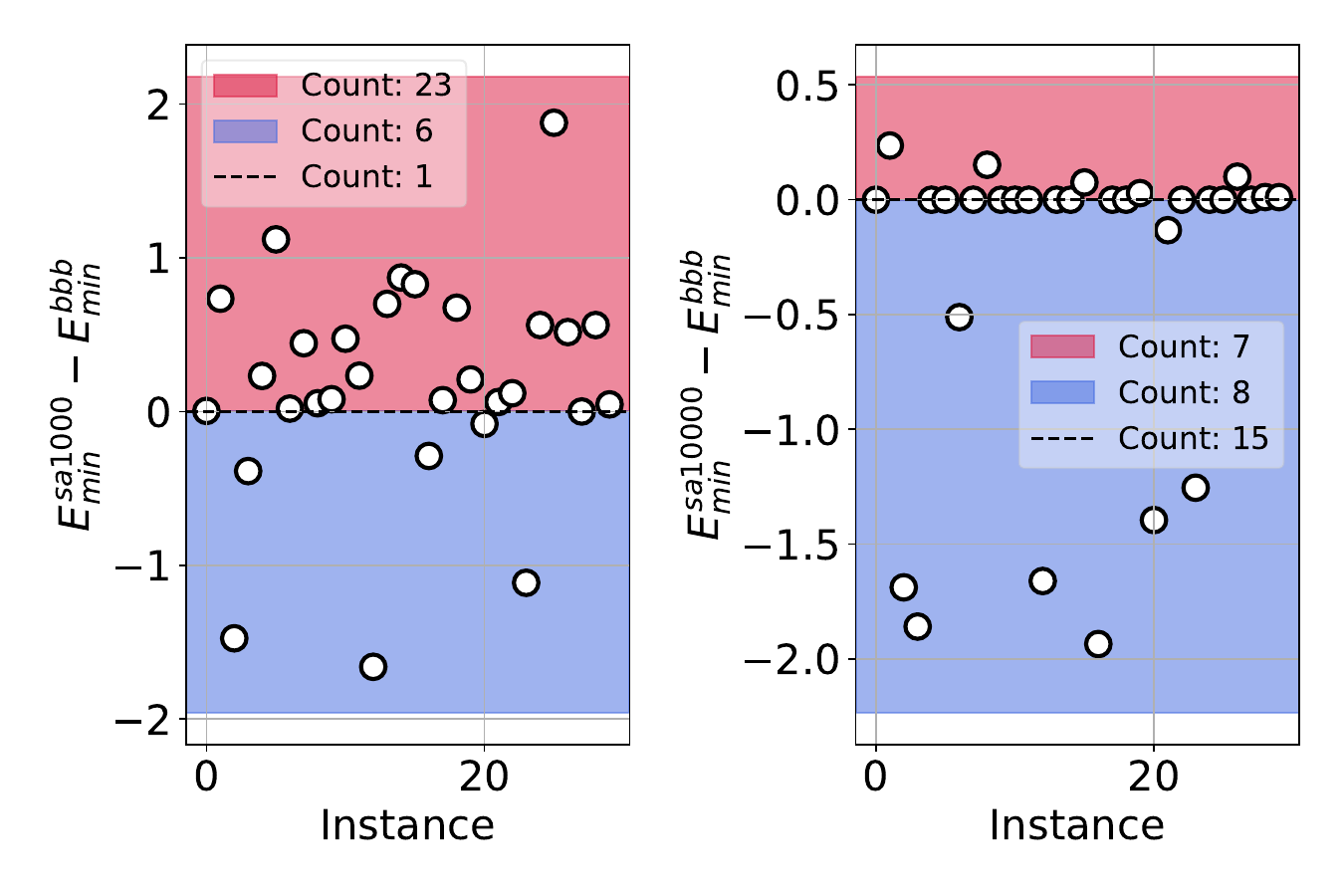}
    \caption{Energy differences for 30 sparse, 156-qubit instances. Each point represents the energy difference between SA (1000 and 10000 sweeps, respectively) and BBB-DCQO. Negative (positive) values indicate lower (larger) energy for SA compared to BBB-DCQO.}
    \label{fig:mps_30inst}
\end{figure}

\section{Conclusion}\label{sec:conclusion}

Building on the binary branch-and-bound algorithm, in this work we propose a purely quantum algorithm for solving efficiently HUBO problems on current quantum hardware. The BBB-DCQO algorithm does not involve any variational method, thus avoiding potential trainability drawbacks such as barren plateaus. Additionally, it does not require any extra qubits for HUBO-to-QUBO mappings, unlike the classical branch-and-bound or quantum annealing. To demonstrate the superior performance of our proposal against well-established methods, we started by comparing the quality of the solutions obtained by BBB-DCQO on an ideal simulator with SA over a set of sparse 156-qubit HUBO instances. For a fair comparison, we restricted both algorithms to consider the same number of function evaluations, namely an equal number of measurements and spin-flips for BBB-DCQO and SA, respectively. To experimentally validate our findings, we use~\texttt{IBM} quantum processors for 100-qubit denser HUBO instances. In this case, we were able to reach the optimal solution with a considerably lower number of function evaluations compared to SA. Consequently, the BBB-DCQO algorithm emerges as a suitable method for solving denser HUBO problems on current quantum platforms, improving upon the previously proposed BF-DCQO algorithm. This paves the path towards large-scale optimization, which could showcase an advantage over well-established classical solvers. Since the method has shown the ability to find high-quality solutions already in the early layers of the binary tree, this work suggests that BBB-DCQO is a promising alternative to classical branch-and-bound algorithms. This encourages future research aimed at developing strategies to verify the optimality of the solutions without relying on classical algorithms.

\begin{acknowledgments}
    We thank Archismita Dalal, Pranav Chandarana, and Murilo Henrique de Oliveira for their valuable discussions and support. We acknowledge the use of IBM Quantum services for this work. The views expressed are those of the authors and do not reflect the official policy or position of IBM or the IBM Quantum team. 
\end{acknowledgments}

\vspace{0.5cm}

\appendix
\section{Circuit decomposition on IBM hardware}\label{sec:ibm}

A key aspect of preparing and running quantum circuits on hardware is to transpile the required quantum operations into the available ones in the targeted platform. These typically consist of a universal gate set containing several one-qubit gates and a single two-qubit entangling gate. In our experiments, we used \textsc{ibm\_marrakesh}, gate-based quantum computer composed of 156 superconducting qubits under a heavy-hexagonal coupling map. Its native gate set is composed of
\begin{equation}
    X=\begin{bmatrix}0&1\\1&0\end{bmatrix},\text{ }\sqrt{X}=\frac{1}{2}\begin{bmatrix}1+i & 1-i\\ 1-i& 1+i\end{bmatrix},\text{ } R_z(\theta)=e^{-i\theta \sigma^z/2},
\end{equation}
with $\text{CZ}=\diag(1,1,1,-1)$ as an entangling gate. Apart from them, IBM has recently introduced in their Heron-based processors the fractional gates $R_{zz}(\theta)=e^{-i\theta \sigma^z_0\sigma^z_1/2}$ (with $0<\theta\le\pi/2$) and $R_x(\theta)=e^{-i\theta \sigma^x/2}$~\cite{frac}.

\section{Post-processing}
\label{appendix:post_processing}

The greedy local search algorithm (Algorithm~\ref{alg:greedy}) acts as a zero-temperature SA scheme that can be executed over a fixed amount of sweeps. It aims to correct detrimental bit flips, hence locally improving the quality of the solution by greedily accepting only cost-reducing single-bit changes. Though limited by its tendency to get trapped in local minima, it is lightweight and effective for local refinement, making it a valuable post-processing step when combined with more exploratory or stochastic optimization methods.
\begin{figure}[htb] 
\vspace{-2.7mm}
\begin{minipage}[t]{\columnwidth}%
\begin{algorithm}[H]%
   \begin{algorithmic}[1]%
    \State \textbf{Input:} bitstring $z$, cost function $F(z)$
    \For{$z_i$ in \texttt{shuffle}($z$)} \Comment{shuffle indices to be flipped}
        \State $z' = \texttt{flip}(z, z_i)$ \Comment{flip $i$th spin value}
        \If{ $F(z') < F(z)$}
            \State $z \leftarrow z'$
        \EndIf
    \EndFor
    \State\textbf{Output:} $z$.
   \end{algorithmic}%
\caption{Greedy local search algorithm.}\label{alg:greedy}%
\end{algorithm}%
\end{minipage}%
\end{figure}%

\bibliography{reference.bib}

\begin{thebibliography}{49}%
\makeatletter
\providecommand \@ifxundefined [1]{%
 \@ifx{#1\undefined}
}%
\providecommand \@ifnum [1]{%
 \ifnum #1\expandafter \@firstoftwo
 \else \expandafter \@secondoftwo
 \fi
}%
\providecommand \@ifx [1]{%
 \ifx #1\expandafter \@firstoftwo
 \else \expandafter \@secondoftwo
 \fi
}%
\providecommand \natexlab [1]{#1}%
\providecommand \enquote  [1]{``#1''}%
\providecommand \bibnamefont  [1]{#1}%
\providecommand \bibfnamefont [1]{#1}%
\providecommand \citenamefont [1]{#1}%
\providecommand \href@noop [0]{\@secondoftwo}%
\providecommand \href [0]{\begingroup \@sanitize@url \@href}%
\providecommand \@href[1]{\@@startlink{#1}\@@href}%
\providecommand \@@href[1]{\endgroup#1\@@endlink}%
\providecommand \@sanitize@url [0]{\catcode `\\12\catcode `\$12\catcode `\&12\catcode `\#12\catcode `\^12\catcode `\_12\catcode `\%12\relax}%
\providecommand \@@startlink[1]{}%
\providecommand \@@endlink[0]{}%
\providecommand \url  [0]{\begingroup\@sanitize@url \@url }%
\providecommand \@url [1]{\endgroup\@href {#1}{\urlprefix }}%
\providecommand \urlprefix  [0]{URL }%
\providecommand \Eprint [0]{\href }%
\providecommand \doibase [0]{https://doi.org/}%
\providecommand \selectlanguage [0]{\@gobble}%
\providecommand \bibinfo  [0]{\@secondoftwo}%
\providecommand \bibfield  [0]{\@secondoftwo}%
\providecommand \translation [1]{[#1]}%
\providecommand \BibitemOpen [0]{}%
\providecommand \bibitemStop [0]{}%
\providecommand \bibitemNoStop [0]{.\EOS\space}%
\providecommand \EOS [0]{\spacefactor3000\relax}%
\providecommand \BibitemShut  [1]{\csname bibitem#1\endcsname}%
\let\auto@bib@innerbib\@empty
\bibitem [{\citenamefont {Floudas}\ \emph {et~al.}(2005)\citenamefont {Floudas}, \citenamefont {Akrotirianakis}, \citenamefont {Caratzoulas}, \citenamefont {Meyer},\ and\ \citenamefont {Kallrath}}]{nonlinearprog}%
  \BibitemOpen
  \bibfield  {author} {\bibinfo {author} {\bibfnamefont {C.~A.}\ \bibnamefont {Floudas}}, \bibinfo {author} {\bibfnamefont {I.~G.}\ \bibnamefont {Akrotirianakis}}, \bibinfo {author} {\bibfnamefont {S.}~\bibnamefont {Caratzoulas}}, \bibinfo {author} {\bibfnamefont {C.~A.}\ \bibnamefont {Meyer}},\ and\ \bibinfo {author} {\bibfnamefont {J.}~\bibnamefont {Kallrath}},\ }\href {https://doi.org/10.1016/j.compchemeng.2005.02.006} {\bibfield  {journal} {\bibinfo  {journal} {Computers \& Chemical Engineering}\ }\textbf {\bibinfo {volume} {29}},\ \bibinfo {pages} {1185} (\bibinfo {year} {2005})}\BibitemShut {NoStop}%
\bibitem [{\citenamefont {Burer}\ and\ \citenamefont {Letchford}(2012)}]{surveynonconvex}%
  \BibitemOpen
  \bibfield  {author} {\bibinfo {author} {\bibfnamefont {S.}~\bibnamefont {Burer}}\ and\ \bibinfo {author} {\bibfnamefont {A.~N.}\ \bibnamefont {Letchford}},\ }\href {https://doi.org/10.1016/j.sorms.2012.08.001} {\bibfield  {journal} {\bibinfo  {journal} {Surveys in Operations Research and Management Science}\ }\textbf {\bibinfo {volume} {17}},\ \bibinfo {pages} {97} (\bibinfo {year} {2012})}\BibitemShut {NoStop}%
\bibitem [{\citenamefont {Xu}\ and\ \citenamefont {Liberti}(2024)}]{xu2024relaxationschallenges}%
  \BibitemOpen
  \bibfield  {author} {\bibinfo {author} {\bibfnamefont {L.}~\bibnamefont {Xu}}\ and\ \bibinfo {author} {\bibfnamefont {L.}~\bibnamefont {Liberti}},\ }\href {https://arxiv.org/abs/2405.13447} {\bibinfo {title} {Relaxations for binary polynomial optimization via signed certificates}} (\bibinfo {year} {2024}),\ \Eprint {https://arxiv.org/abs/2405.13447} {arXiv:2405.13447 [math.OC]} \BibitemShut {NoStop}%
\bibitem [{\citenamefont {Danilova}\ \emph {et~al.}(2022)\citenamefont {Danilova}, \citenamefont {Dvurechensky}, \citenamefont {Gasnikov}, \citenamefont {Gorbunov}, \citenamefont {Guminov}, \citenamefont {Kamzolov},\ and\ \citenamefont {Shibaev}}]{danilova2022nonconvexchallenges}%
  \BibitemOpen
  \bibfield  {author} {\bibinfo {author} {\bibfnamefont {M.}~\bibnamefont {Danilova}}, \bibinfo {author} {\bibfnamefont {P.}~\bibnamefont {Dvurechensky}}, \bibinfo {author} {\bibfnamefont {A.}~\bibnamefont {Gasnikov}}, \bibinfo {author} {\bibfnamefont {E.}~\bibnamefont {Gorbunov}}, \bibinfo {author} {\bibfnamefont {S.}~\bibnamefont {Guminov}}, \bibinfo {author} {\bibfnamefont {D.}~\bibnamefont {Kamzolov}},\ and\ \bibinfo {author} {\bibfnamefont {I.}~\bibnamefont {Shibaev}},\ }in\ \href {https://doi.org/10.1007/978-3-031-00832-0_3} {\emph {\bibinfo {booktitle} {{High-Dimensional Optimization and Probability: With a View Towards Data Science}}}},\ \bibinfo {series} {Springer Optimization and Its Applications}, Vol.\ \bibinfo {volume} {191}\ (\bibinfo  {publisher} {Springer},\ \bibinfo {year} {2022})\ pp.\ \bibinfo {pages} {79--163}\BibitemShut {NoStop}%
\bibitem [{\citenamefont {Arjevani}\ \emph {et~al.}(2020)\citenamefont {Arjevani}, \citenamefont {Carmon}, \citenamefont {Duchi}, \citenamefont {Foster}, \citenamefont {Sekhari},\ and\ \citenamefont {Sridharan}}]{arjevani2020nonconvexchallenges}%
  \BibitemOpen
  \bibfield  {author} {\bibinfo {author} {\bibfnamefont {Y.}~\bibnamefont {Arjevani}}, \bibinfo {author} {\bibfnamefont {Y.}~\bibnamefont {Carmon}}, \bibinfo {author} {\bibfnamefont {J.~C.}\ \bibnamefont {Duchi}}, \bibinfo {author} {\bibfnamefont {D.~J.}\ \bibnamefont {Foster}}, \bibinfo {author} {\bibfnamefont {A.}~\bibnamefont {Sekhari}},\ and\ \bibinfo {author} {\bibfnamefont {K.}~\bibnamefont {Sridharan}},\ }in\ \href {https://proceedings.mlr.press/v125/arjevani20a.html} {\emph {\bibinfo {booktitle} {{Proceedings of the 33rd Conference on Learning Theory}}}},\ \bibinfo {series} {Proceedings of Machine Learning Research}, Vol.\ \bibinfo {volume} {125}\ (\bibinfo  {publisher} {PMLR},\ \bibinfo {year} {2020})\ pp.\ \bibinfo {pages} {242--299}\BibitemShut {NoStop}%
\bibitem [{\citenamefont {Bertsekas}\ \emph {et~al.}(2003)\citenamefont {Bertsekas}, \citenamefont {Nedic},\ and\ \citenamefont {Ozdaglar}}]{bertsekas2003convex}%
  \BibitemOpen
  \bibfield  {author} {\bibinfo {author} {\bibfnamefont {D.~P.}\ \bibnamefont {Bertsekas}}, \bibinfo {author} {\bibfnamefont {A.}~\bibnamefont {Nedic}},\ and\ \bibinfo {author} {\bibfnamefont {A.~E.}\ \bibnamefont {Ozdaglar}},\ }\href {https://www.athenasc.com/convexity.html} {\emph {\bibinfo {title} {Convex Analysis and Optimization}}}\ (\bibinfo  {publisher} {Athena Scientific},\ \bibinfo {address} {Belmont, MA},\ \bibinfo {year} {2003})\BibitemShut {NoStop}%
\bibitem [{\citenamefont {Cplex}(2009)}]{cplex}%
  \BibitemOpen
  \bibfield  {author} {\bibinfo {author} {\bibfnamefont {I.~I.}\ \bibnamefont {Cplex}},\ }\href@noop {} {\bibfield  {journal} {\bibinfo  {journal} {International Business Machines Corporation}\ }\textbf {\bibinfo {volume} {46}},\ \bibinfo {pages} {157} (\bibinfo {year} {2009})}\BibitemShut {NoStop}%
\bibitem [{\citenamefont {{Gurobi Optimization, LLC}}(2023)}]{gurobi}%
  \BibitemOpen
  \bibfield  {author} {\bibinfo {author} {\bibnamefont {{Gurobi Optimization, LLC}}},\ }\href {https://www.gurobi.com} {\bibinfo {title} {{Gurobi Optimizer Reference Manual}}} (\bibinfo {year} {2023})\BibitemShut {NoStop}%
\bibitem [{\citenamefont {Lucas}(2014)}]{lucas2014ising}%
  \BibitemOpen
  \bibfield  {author} {\bibinfo {author} {\bibfnamefont {A.}~\bibnamefont {Lucas}},\ }\href {https://doi.org/10.3389/fphy.2014.00005} {\bibfield  {journal} {\bibinfo  {journal} {Frontiers in Physics}\ }\textbf {\bibinfo {volume} {2}},\ \bibinfo {pages} {5} (\bibinfo {year} {2014})}\BibitemShut {NoStop}%
\bibitem [{\citenamefont {Kirkpatrick}\ \emph {et~al.}(1983{\natexlab{a}})\citenamefont {Kirkpatrick}, \citenamefont {Gelatt},\ and\ \citenamefont {Vecchi}}]{sim_annealing}%
  \BibitemOpen
  \bibfield  {author} {\bibinfo {author} {\bibfnamefont {S.}~\bibnamefont {Kirkpatrick}}, \bibinfo {author} {\bibfnamefont {C.~D.}\ \bibnamefont {Gelatt}},\ and\ \bibinfo {author} {\bibfnamefont {M.~P.}\ \bibnamefont {Vecchi}},\ }\href {https://doi.org/10.1126/science.220.4598.671} {\bibfield  {journal} {\bibinfo  {journal} {Science}\ }\textbf {\bibinfo {volume} {220}},\ \bibinfo {pages} {671} (\bibinfo {year} {1983}{\natexlab{a}})}\BibitemShut {NoStop}%
\bibitem [{\citenamefont {Glover}\ \emph {et~al.}(1993)\citenamefont {Glover}, \citenamefont {Taillard},\ and\ \citenamefont {de~Werra}}]{tabu}%
  \BibitemOpen
  \bibfield  {author} {\bibinfo {author} {\bibfnamefont {F.}~\bibnamefont {Glover}}, \bibinfo {author} {\bibfnamefont {E.}~\bibnamefont {Taillard}},\ and\ \bibinfo {author} {\bibfnamefont {D.}~\bibnamefont {de~Werra}},\ }\href {https://doi.org/10.1007/BF02078647} {\bibfield  {journal} {\bibinfo  {journal} {Annals of Operations Research}\ }\textbf {\bibinfo {volume} {41}},\ \bibinfo {pages} {3} (\bibinfo {year} {1993})}\BibitemShut {NoStop}%
\bibitem [{\citenamefont {Crisanti}\ and\ \citenamefont {Sommers}(1992)}]{crisanti1992spherical}%
  \BibitemOpen
  \bibfield  {author} {\bibinfo {author} {\bibfnamefont {A.}~\bibnamefont {Crisanti}}\ and\ \bibinfo {author} {\bibfnamefont {H.~J.}\ \bibnamefont {Sommers}},\ }\href {https://doi.org/10.1007/BF01309287} {\bibfield  {journal} {\bibinfo  {journal} {Zeitschrift für Physik B Condensed Matter}\ }\textbf {\bibinfo {volume} {87}},\ \bibinfo {pages} {341–354} (\bibinfo {year} {1992})}\BibitemShut {NoStop}%
\bibitem [{\citenamefont {Pelofske}\ \emph {et~al.}(2023)\citenamefont {Pelofske}, \citenamefont {B{\"a}rtschi},\ and\ \citenamefont {Eidenbenz}}]{pelofske2023quantum}%
  \BibitemOpen
  \bibfield  {author} {\bibinfo {author} {\bibfnamefont {E.}~\bibnamefont {Pelofske}}, \bibinfo {author} {\bibfnamefont {A.}~\bibnamefont {B{\"a}rtschi}},\ and\ \bibinfo {author} {\bibfnamefont {S.}~\bibnamefont {Eidenbenz}},\ }in\ \href {https://doi.org/10.1007/978-3-031-32041-5_13} {\emph {\bibinfo {booktitle} {{High Performance Computing: 38th International Conference, ISC High Performance 2023, Hamburg, Germany, May 21--25, 2023, Proceedings}}}},\ \bibinfo {series} {Lecture Notes in Computer Science}, Vol.\ \bibinfo {volume} {13944}\ (\bibinfo  {publisher} {Springer Nature Switzerland},\ \bibinfo {address} {Cham},\ \bibinfo {year} {2023})\ pp.\ \bibinfo {pages} {240--258}\BibitemShut {NoStop}%
\bibitem [{\citenamefont {Pelofske}\ \emph {et~al.}(2024)\citenamefont {Pelofske}, \citenamefont {Bärtschi},\ and\ \citenamefont {Eidenbenz}}]{pelofske2024short-depth}%
  \BibitemOpen
  \bibfield  {author} {\bibinfo {author} {\bibfnamefont {E.}~\bibnamefont {Pelofske}}, \bibinfo {author} {\bibfnamefont {A.}~\bibnamefont {Bärtschi}},\ and\ \bibinfo {author} {\bibfnamefont {S.}~\bibnamefont {Eidenbenz}},\ }\href {https://doi.org/10.1038/s41534-024-00825-w} {\bibfield  {journal} {\bibinfo  {journal} {npj Quantum Information}\ }\textbf {\bibinfo {volume} {10}},\ \bibinfo {pages} {1–19} (\bibinfo {year} {2024})}\BibitemShut {NoStop}%
\bibitem [{\citenamefont {Robert}\ \emph {et~al.}(2021)\citenamefont {Robert}, \citenamefont {Barkoutsos}, \citenamefont {Woerner},\ and\ \citenamefont {Tavernelli}}]{robert2021resource}%
  \BibitemOpen
  \bibfield  {author} {\bibinfo {author} {\bibfnamefont {A.}~\bibnamefont {Robert}}, \bibinfo {author} {\bibfnamefont {P.~K.}\ \bibnamefont {Barkoutsos}}, \bibinfo {author} {\bibfnamefont {S.}~\bibnamefont {Woerner}},\ and\ \bibinfo {author} {\bibfnamefont {I.}~\bibnamefont {Tavernelli}},\ }\href {https://doi.org/10.1038/s41534-021-00368-4} {\bibfield  {journal} {\bibinfo  {journal} {npj Quantum Information}\ }\textbf {\bibinfo {volume} {7}},\ \bibinfo {pages} {38} (\bibinfo {year} {2021})}\BibitemShut {NoStop}%
\bibitem [{\citenamefont {Boulebnane}\ and\ \citenamefont {Montanaro}(2024)}]{boulebnane2022solving}%
  \BibitemOpen
  \bibfield  {author} {\bibinfo {author} {\bibfnamefont {S.}~\bibnamefont {Boulebnane}}\ and\ \bibinfo {author} {\bibfnamefont {A.}~\bibnamefont {Montanaro}},\ }\href {https://doi.org/10.1103/PRXQuantum.5.030348} {\bibfield  {journal} {\bibinfo  {journal} {PRX Quantum}\ }\textbf {\bibinfo {volume} {5}},\ \bibinfo {pages} {030348} (\bibinfo {year} {2024})}\BibitemShut {NoStop}%
\bibitem [{\citenamefont {Abbas}\ \emph {et~al.}(2024)\citenamefont {Abbas}, \citenamefont {Ambainis}, \citenamefont {Augustino}, \citenamefont {B{\"a}rtschi}, \citenamefont {Buhrman}, \citenamefont {Coffrin}, \citenamefont {Cortiana}, \citenamefont {Dunjko}, \citenamefont {Egger}, \citenamefont {Elmegreen} \emph {et~al.}}]{abbas2024challenges}%
  \BibitemOpen
  \bibfield  {author} {\bibinfo {author} {\bibfnamefont {A.}~\bibnamefont {Abbas}}, \bibinfo {author} {\bibfnamefont {A.}~\bibnamefont {Ambainis}}, \bibinfo {author} {\bibfnamefont {B.}~\bibnamefont {Augustino}}, \bibinfo {author} {\bibfnamefont {A.}~\bibnamefont {B{\"a}rtschi}}, \bibinfo {author} {\bibfnamefont {H.}~\bibnamefont {Buhrman}}, \bibinfo {author} {\bibfnamefont {C.}~\bibnamefont {Coffrin}}, \bibinfo {author} {\bibfnamefont {G.}~\bibnamefont {Cortiana}}, \bibinfo {author} {\bibfnamefont {V.}~\bibnamefont {Dunjko}}, \bibinfo {author} {\bibfnamefont {D.~J.}\ \bibnamefont {Egger}}, \bibinfo {author} {\bibfnamefont {B.~G.}\ \bibnamefont {Elmegreen}}, \emph {et~al.},\ }\href {https://www.nature.com/articles/s42254-024-00770-9} {\bibfield  {journal} {\bibinfo  {journal} {Nature Reviews Physics}\ ,\ \bibinfo {pages} {1}} (\bibinfo {year} {2024})}\BibitemShut {NoStop}%
\bibitem [{\citenamefont {Koch}\ \emph {et~al.}(2025)\citenamefont {Koch}, \citenamefont {Neira}, \citenamefont {Chen}, \citenamefont {Cortiana}, \citenamefont {Egger}, \citenamefont {Heese}, \citenamefont {Hegade}, \citenamefont {Cadavid}, \citenamefont {Huang}, \citenamefont {Itoko} \emph {et~al.}}]{koch2025quantum}%
  \BibitemOpen
  \bibfield  {author} {\bibinfo {author} {\bibfnamefont {T.}~\bibnamefont {Koch}}, \bibinfo {author} {\bibfnamefont {D.~E.~B.}\ \bibnamefont {Neira}}, \bibinfo {author} {\bibfnamefont {Y.}~\bibnamefont {Chen}}, \bibinfo {author} {\bibfnamefont {G.}~\bibnamefont {Cortiana}}, \bibinfo {author} {\bibfnamefont {D.~J.}\ \bibnamefont {Egger}}, \bibinfo {author} {\bibfnamefont {R.}~\bibnamefont {Heese}}, \bibinfo {author} {\bibfnamefont {N.~N.}\ \bibnamefont {Hegade}}, \bibinfo {author} {\bibfnamefont {A.~G.}\ \bibnamefont {Cadavid}}, \bibinfo {author} {\bibfnamefont {R.}~\bibnamefont {Huang}}, \bibinfo {author} {\bibfnamefont {T.}~\bibnamefont {Itoko}}, \emph {et~al.},\ }\href {https://arxiv.org/abs/2504.03832} {\bibinfo {title} {{Quantum Optimization Benchmark Library -- The Intractable Decathlon}}} (\bibinfo {year} {2025}),\ \Eprint {https://arxiv.org/abs/2504.03832} {arXiv:2504.03832 [quant-ph]} \BibitemShut {NoStop}%
\bibitem [{\citenamefont {Kotil}\ \emph {et~al.}(2025)\citenamefont {Kotil}, \citenamefont {Pelofske}, \citenamefont {Riedmüller}, \citenamefont {Egger}, \citenamefont {Eidenbenz}, \citenamefont {Koch},\ and\ \citenamefont {Woerner}}]{kotil2025quantum}%
  \BibitemOpen
  \bibfield  {author} {\bibinfo {author} {\bibfnamefont {A.}~\bibnamefont {Kotil}}, \bibinfo {author} {\bibfnamefont {E.}~\bibnamefont {Pelofske}}, \bibinfo {author} {\bibfnamefont {S.}~\bibnamefont {Riedmüller}}, \bibinfo {author} {\bibfnamefont {D.~J.}\ \bibnamefont {Egger}}, \bibinfo {author} {\bibfnamefont {S.}~\bibnamefont {Eidenbenz}}, \bibinfo {author} {\bibfnamefont {T.}~\bibnamefont {Koch}},\ and\ \bibinfo {author} {\bibfnamefont {S.}~\bibnamefont {Woerner}},\ }\href {https://arxiv.org/abs/2503.22797} {\bibinfo {title} {{Quantum Approximate Multi-Objective Optimization}}} (\bibinfo {year} {2025}),\ \Eprint {https://arxiv.org/abs/2503.22797} {arXiv:2503.22797 [quant-ph]} \BibitemShut {NoStop}%
\bibitem [{\citenamefont {Farhi}\ \emph {et~al.}(2014)\citenamefont {Farhi}, \citenamefont {Goldstone},\ and\ \citenamefont {Gutmann}}]{farhi2014quantum}%
  \BibitemOpen
  \bibfield  {author} {\bibinfo {author} {\bibfnamefont {E.}~\bibnamefont {Farhi}}, \bibinfo {author} {\bibfnamefont {J.}~\bibnamefont {Goldstone}},\ and\ \bibinfo {author} {\bibfnamefont {S.}~\bibnamefont {Gutmann}},\ }\href {https://arxiv.org/abs/1411.4028} {\bibinfo {title} {{A Quantum Approximate Optimization Algorithm}}} (\bibinfo {year} {2014}),\ \Eprint {https://arxiv.org/abs/1411.4028} {arXiv:1411.4028 [quant-ph]} \BibitemShut {NoStop}%
\bibitem [{\citenamefont {Cerezo}\ \emph {et~al.}(2024)\citenamefont {Cerezo}, \citenamefont {Larocca}, \citenamefont {García-Martín}, \citenamefont {Diaz}, \citenamefont {Braccia}, \citenamefont {Fontana}, \citenamefont {Rudolph}, \citenamefont {Bermejo}, \citenamefont {Ijaz}, \citenamefont {Thanasilp}, \citenamefont {Anschuetz},\ and\ \citenamefont {Holmes}}]{cerezo2024doesprovableabsencebarren}%
  \BibitemOpen
  \bibfield  {author} {\bibinfo {author} {\bibfnamefont {M.}~\bibnamefont {Cerezo}}, \bibinfo {author} {\bibfnamefont {M.}~\bibnamefont {Larocca}}, \bibinfo {author} {\bibfnamefont {D.}~\bibnamefont {García-Martín}}, \bibinfo {author} {\bibfnamefont {N.~L.}\ \bibnamefont {Diaz}}, \bibinfo {author} {\bibfnamefont {P.}~\bibnamefont {Braccia}}, \bibinfo {author} {\bibfnamefont {E.}~\bibnamefont {Fontana}}, \bibinfo {author} {\bibfnamefont {M.~S.}\ \bibnamefont {Rudolph}}, \bibinfo {author} {\bibfnamefont {P.}~\bibnamefont {Bermejo}}, \bibinfo {author} {\bibfnamefont {A.}~\bibnamefont {Ijaz}}, \bibinfo {author} {\bibfnamefont {S.}~\bibnamefont {Thanasilp}}, \bibinfo {author} {\bibfnamefont {E.~R.}\ \bibnamefont {Anschuetz}},\ and\ \bibinfo {author} {\bibfnamefont {Z.}~\bibnamefont {Holmes}},\ }\href {https://arxiv.org/abs/2312.09121} {\bibinfo {title} {{Does provable absence of barren plateaus imply classical simulability? Or, why we need to rethink variational quantum computing}}} (\bibinfo {year} {2024}),\ \Eprint {https://arxiv.org/abs/2312.09121} {arXiv:2312.09121 [quant-ph]} \BibitemShut {NoStop}%
\bibitem [{\citenamefont {Larocca}\ \emph {et~al.}(2025)\citenamefont {Larocca}, \citenamefont {Thanasilp}, \citenamefont {Wang}, \citenamefont {Sharma}, \citenamefont {Biamonte}, \citenamefont {Coles}, \citenamefont {Cincio}, \citenamefont {McClean}, \citenamefont {Holmes},\ and\ \citenamefont {Cerezo}}]{larocca2024reviewbarrenplateausvariational}%
  \BibitemOpen
  \bibfield  {author} {\bibinfo {author} {\bibfnamefont {M.}~\bibnamefont {Larocca}}, \bibinfo {author} {\bibfnamefont {S.}~\bibnamefont {Thanasilp}}, \bibinfo {author} {\bibfnamefont {S.}~\bibnamefont {Wang}}, \bibinfo {author} {\bibfnamefont {K.}~\bibnamefont {Sharma}}, \bibinfo {author} {\bibfnamefont {J.}~\bibnamefont {Biamonte}}, \bibinfo {author} {\bibfnamefont {P.~J.}\ \bibnamefont {Coles}}, \bibinfo {author} {\bibfnamefont {L.}~\bibnamefont {Cincio}}, \bibinfo {author} {\bibfnamefont {J.~R.}\ \bibnamefont {McClean}}, \bibinfo {author} {\bibfnamefont {Z.}~\bibnamefont {Holmes}},\ and\ \bibinfo {author} {\bibfnamefont {M.}~\bibnamefont {Cerezo}},\ }\href {https://doi.org/10.1038/s42254-025-00813-9} {\bibfield  {journal} {\bibinfo  {journal} {Nature Reviews Physics}\ }\textbf {\bibinfo {volume} {7}},\ \bibinfo {pages} {174–189} (\bibinfo {year} {2025})}\BibitemShut {NoStop}%
\bibitem [{\citenamefont {Kolodrubetz}\ \emph {et~al.}(2017)\citenamefont {Kolodrubetz}, \citenamefont {Sels}, \citenamefont {Mehta},\ and\ \citenamefont {Polkovnikov}}]{kolodrubetz2017geometry}%
  \BibitemOpen
  \bibfield  {author} {\bibinfo {author} {\bibfnamefont {M.}~\bibnamefont {Kolodrubetz}}, \bibinfo {author} {\bibfnamefont {D.}~\bibnamefont {Sels}}, \bibinfo {author} {\bibfnamefont {P.}~\bibnamefont {Mehta}},\ and\ \bibinfo {author} {\bibfnamefont {A.}~\bibnamefont {Polkovnikov}},\ }\href {https://www.sciencedirect.com/science/article/abs/pii/S0370157317301989} {\bibfield  {journal} {\bibinfo  {journal} {Physics Reports}\ }\textbf {\bibinfo {volume} {697}},\ \bibinfo {pages} {1} (\bibinfo {year} {2017})}\BibitemShut {NoStop}%
\bibitem [{\citenamefont {Sels}\ and\ \citenamefont {Polkovnikov}(2017)}]{sels2017minimizing}%
  \BibitemOpen
  \bibfield  {author} {\bibinfo {author} {\bibfnamefont {D.}~\bibnamefont {Sels}}\ and\ \bibinfo {author} {\bibfnamefont {A.}~\bibnamefont {Polkovnikov}},\ }\href {https://www.pnas.org/doi/full/10.1073/pnas.1619826114} {\bibfield  {journal} {\bibinfo  {journal} {Proceedings of the National Academy of Sciences}\ }\textbf {\bibinfo {volume} {114}},\ \bibinfo {pages} {E3909} (\bibinfo {year} {2017})}\BibitemShut {NoStop}%
\bibitem [{\citenamefont {Claeys}\ \emph {et~al.}(2019)\citenamefont {Claeys}, \citenamefont {Pandey}, \citenamefont {Sels},\ and\ \citenamefont {Polkovnikov}}]{claeys2019floquet}%
  \BibitemOpen
  \bibfield  {author} {\bibinfo {author} {\bibfnamefont {P.~W.}\ \bibnamefont {Claeys}}, \bibinfo {author} {\bibfnamefont {M.}~\bibnamefont {Pandey}}, \bibinfo {author} {\bibfnamefont {D.}~\bibnamefont {Sels}},\ and\ \bibinfo {author} {\bibfnamefont {A.}~\bibnamefont {Polkovnikov}},\ }\href {https://journals.aps.org/prl/abstract/10.1103/PhysRevLett.123.090602} {\bibfield  {journal} {\bibinfo  {journal} {Phys. Rev. Lett.}\ }\textbf {\bibinfo {volume} {123}},\ \bibinfo {pages} {090602} (\bibinfo {year} {2019})}\BibitemShut {NoStop}%
\bibitem [{\citenamefont {Hatomura}\ and\ \citenamefont {Takahashi}(2021)}]{hatomura2021controlling}%
  \BibitemOpen
  \bibfield  {author} {\bibinfo {author} {\bibfnamefont {T.}~\bibnamefont {Hatomura}}\ and\ \bibinfo {author} {\bibfnamefont {K.}~\bibnamefont {Takahashi}},\ }\href {https://journals.aps.org/pra/abstract/10.1103/PhysRevA.103.012220} {\bibfield  {journal} {\bibinfo  {journal} {Phys. Rev. A}\ }\textbf {\bibinfo {volume} {103}},\ \bibinfo {pages} {012220} (\bibinfo {year} {2021})}\BibitemShut {NoStop}%
\bibitem [{\citenamefont {Takahashi}\ and\ \citenamefont {del Campo}(2024)}]{takahashi2024shortcuts}%
  \BibitemOpen
  \bibfield  {author} {\bibinfo {author} {\bibfnamefont {K.}~\bibnamefont {Takahashi}}\ and\ \bibinfo {author} {\bibfnamefont {A.}~\bibnamefont {del Campo}},\ }\href {https://journals.aps.org/prx/abstract/10.1103/PhysRevX.14.011032} {\bibfield  {journal} {\bibinfo  {journal} {Phys. Rev. X}\ }\textbf {\bibinfo {volume} {14}},\ \bibinfo {pages} {011032} (\bibinfo {year} {2024})}\BibitemShut {NoStop}%
\bibitem [{\citenamefont {Del~Campo}(2013)}]{del2013shortcuts}%
  \BibitemOpen
  \bibfield  {author} {\bibinfo {author} {\bibfnamefont {A.}~\bibnamefont {Del~Campo}},\ }\href {https://journals.aps.org/prl/abstract/10.1103/PhysRevLett.111.100502} {\bibfield  {journal} {\bibinfo  {journal} {Phys. Rev. Lett.}\ }\textbf {\bibinfo {volume} {111}},\ \bibinfo {pages} {100502} (\bibinfo {year} {2013})}\BibitemShut {NoStop}%
\bibitem [{\citenamefont {Hegade}\ \emph {et~al.}(2021)\citenamefont {Hegade}, \citenamefont {Paul}, \citenamefont {Ding}, \citenamefont {Sanz}, \citenamefont {Albarr{\'a}n-Arriagada}, \citenamefont {Solano},\ and\ \citenamefont {Chen}}]{hegade2021shortcuts}%
  \BibitemOpen
  \bibfield  {author} {\bibinfo {author} {\bibfnamefont {N.~N.}\ \bibnamefont {Hegade}}, \bibinfo {author} {\bibfnamefont {K.}~\bibnamefont {Paul}}, \bibinfo {author} {\bibfnamefont {Y.}~\bibnamefont {Ding}}, \bibinfo {author} {\bibfnamefont {M.}~\bibnamefont {Sanz}}, \bibinfo {author} {\bibfnamefont {F.}~\bibnamefont {Albarr{\'a}n-Arriagada}}, \bibinfo {author} {\bibfnamefont {E.}~\bibnamefont {Solano}},\ and\ \bibinfo {author} {\bibfnamefont {X.}~\bibnamefont {Chen}},\ }\href {https://journals.aps.org/prapplied/abstract/10.1103/PhysRevApplied.15.024038} {\bibfield  {journal} {\bibinfo  {journal} {Phys. Rev. Appl.}\ }\textbf {\bibinfo {volume} {15}},\ \bibinfo {pages} {024038} (\bibinfo {year} {2021})}\BibitemShut {NoStop}%
\bibitem [{\citenamefont {Demirplak}\ and\ \citenamefont {Rice}(2003)}]{demirplak2003adiabatic}%
  \BibitemOpen
  \bibfield  {author} {\bibinfo {author} {\bibfnamefont {M.}~\bibnamefont {Demirplak}}\ and\ \bibinfo {author} {\bibfnamefont {S.~A.}\ \bibnamefont {Rice}},\ }\href {https://pubs.acs.org/doi/10.1021/jp030708a} {\bibfield  {journal} {\bibinfo  {journal} {The Journal of Physical Chemistry A}\ }\textbf {\bibinfo {volume} {107}},\ \bibinfo {pages} {9937} (\bibinfo {year} {2003})}\BibitemShut {NoStop}%
\bibitem [{\citenamefont {Berry}(2009)}]{berry2009transitionless}%
  \BibitemOpen
  \bibfield  {author} {\bibinfo {author} {\bibfnamefont {M.~V.}\ \bibnamefont {Berry}},\ }\href {https://iopscience.iop.org/article/10.1088/1751-8113/42/36/365303} {\bibfield  {journal} {\bibinfo  {journal} {Journal of Physics A: Mathematical and Theoretical}\ }\textbf {\bibinfo {volume} {42}},\ \bibinfo {pages} {365303} (\bibinfo {year} {2009})}\BibitemShut {NoStop}%
\bibitem [{\citenamefont {Hegade}\ \emph {et~al.}(2022)\citenamefont {Hegade}, \citenamefont {Chen},\ and\ \citenamefont {Solano}}]{hegade2022digitized}%
  \BibitemOpen
  \bibfield  {author} {\bibinfo {author} {\bibfnamefont {N.~N.}\ \bibnamefont {Hegade}}, \bibinfo {author} {\bibfnamefont {X.}~\bibnamefont {Chen}},\ and\ \bibinfo {author} {\bibfnamefont {E.}~\bibnamefont {Solano}},\ }\href {https://journals.aps.org/prresearch/abstract/10.1103/PhysRevResearch.4.L042030} {\bibfield  {journal} {\bibinfo  {journal} {Phys. Rev. Res.}\ }\textbf {\bibinfo {volume} {4}},\ \bibinfo {pages} {L042030} (\bibinfo {year} {2022})}\BibitemShut {NoStop}%
\bibitem [{\citenamefont {Cadavid}\ \emph {et~al.}(2025)\citenamefont {Cadavid}, \citenamefont {Dalal}, \citenamefont {Simen}, \citenamefont {Solano},\ and\ \citenamefont {Hegade}}]{cadavid2024bias}%
  \BibitemOpen
  \bibfield  {author} {\bibinfo {author} {\bibfnamefont {A.~G.}\ \bibnamefont {Cadavid}}, \bibinfo {author} {\bibfnamefont {A.}~\bibnamefont {Dalal}}, \bibinfo {author} {\bibfnamefont {A.}~\bibnamefont {Simen}}, \bibinfo {author} {\bibfnamefont {E.}~\bibnamefont {Solano}},\ and\ \bibinfo {author} {\bibfnamefont {N.~N.}\ \bibnamefont {Hegade}},\ }\href {https://doi.org/10.1103/PhysRevResearch.7.L022010} {\bibfield  {journal} {\bibinfo  {journal} {Phys. Rev. Res.}\ }\textbf {\bibinfo {volume} {7}},\ \bibinfo {pages} {L022010} (\bibinfo {year} {2025})}\BibitemShut {NoStop}%
\bibitem [{\citenamefont {Romero}\ \emph {et~al.}(2024{\natexlab{a}})\citenamefont {Romero}, \citenamefont {Visuri}, \citenamefont {Cadavid}, \citenamefont {Solano},\ and\ \citenamefont {Hegade}}]{romero2024bias}%
  \BibitemOpen
  \bibfield  {author} {\bibinfo {author} {\bibfnamefont {S.~V.}\ \bibnamefont {Romero}}, \bibinfo {author} {\bibfnamefont {A.-M.}\ \bibnamefont {Visuri}}, \bibinfo {author} {\bibfnamefont {A.~G.}\ \bibnamefont {Cadavid}}, \bibinfo {author} {\bibfnamefont {E.}~\bibnamefont {Solano}},\ and\ \bibinfo {author} {\bibfnamefont {N.~N.}\ \bibnamefont {Hegade}},\ }\href {https://arxiv.org/abs/2409.04477} {\bibinfo {title} {{Bias-Field Digitized Counterdiabatic Quantum Algorithm for Higher-Order Binary Optimization}}} (\bibinfo {year} {2024}{\natexlab{a}}),\ \Eprint {https://arxiv.org/abs/2409.04477} {arXiv:2409.04477 [quant-ph]} \BibitemShut {NoStop}%
\bibitem [{\citenamefont {{IBM Quantum}}(2025{\natexlab{a}})}]{iskay}%
  \BibitemOpen
  \bibfield  {author} {\bibinfo {author} {\bibnamefont {{IBM Quantum}}},\ }\href@noop {} {\bibinfo {title} {{Iskay Quantum Optimizer - A Qiskit Function by Kipu Quantum}}},\ \bibinfo {howpublished} {\url{https://docs.quantum.ibm.com/guides/kipu-optimization}} (\bibinfo {year} {2025}{\natexlab{a}}),\ \bibinfo {note} {[Online: 14/04/25]}\BibitemShut {NoStop}%
\bibitem [{\citenamefont {Gra{\ss}}(2019)}]{grass2019quantum}%
  \BibitemOpen
  \bibfield  {author} {\bibinfo {author} {\bibfnamefont {T.}~\bibnamefont {Gra{\ss}}},\ }\href {https://journals.aps.org/prl/abstract/10.1103/PhysRevLett.123.120501} {\bibfield  {journal} {\bibinfo  {journal} {Phys. Rev. Lett.}\ }\textbf {\bibinfo {volume} {123}},\ \bibinfo {pages} {120501} (\bibinfo {year} {2019})}\BibitemShut {NoStop}%
\bibitem [{\citenamefont {Grass}(2022)}]{grass2022quantum}%
  \BibitemOpen
  \bibfield  {author} {\bibinfo {author} {\bibfnamefont {T.}~\bibnamefont {Grass}},\ }\href {https://journals.aps.org/prapplied/abstract/10.1103/PhysRevApplied.18.044036} {\bibfield  {journal} {\bibinfo  {journal} {Phys. Rev. Appl.}\ }\textbf {\bibinfo {volume} {18}},\ \bibinfo {pages} {044036} (\bibinfo {year} {2022})}\BibitemShut {NoStop}%
\bibitem [{\citenamefont {Kirkpatrick}\ \emph {et~al.}(1983{\natexlab{b}})\citenamefont {Kirkpatrick}, \citenamefont {Gelatt},\ and\ \citenamefont {Vecchi}}]{kirkpatrick1983optimization}%
  \BibitemOpen
  \bibfield  {author} {\bibinfo {author} {\bibfnamefont {S.}~\bibnamefont {Kirkpatrick}}, \bibinfo {author} {\bibfnamefont {C.~D.}\ \bibnamefont {Gelatt}},\ and\ \bibinfo {author} {\bibfnamefont {M.~P.}\ \bibnamefont {Vecchi}},\ }\href {https://doi.org/10.1126/science.220.4598.671} {\bibfield  {journal} {\bibinfo  {journal} {Science}\ }\textbf {\bibinfo {volume} {220}},\ \bibinfo {pages} {671} (\bibinfo {year} {1983}{\natexlab{b}})}\BibitemShut {NoStop}%
\bibitem [{\citenamefont {{IBM Quantum}}(2025{\natexlab{b}})}]{ibm}%
  \BibitemOpen
  \bibfield  {author} {\bibinfo {author} {\bibnamefont {{IBM Quantum}}},\ }\href@noop {} {}\bibinfo {howpublished} {\url{https://quantum.ibm.com/}} (\bibinfo {year} {2025}{\natexlab{b}})\BibitemShut {NoStop}%
\bibitem [{\citenamefont {{D-Wave Systems}}(2025)}]{dwave}%
  \BibitemOpen
  \bibfield  {author} {\bibinfo {author} {\bibnamefont {{D-Wave Systems}}},\ }\href {https://www.dwavesys.com/} {}\bibinfo {howpublished} {\url{https://www.dwavesys.com/}} (\bibinfo {year} {2025})\BibitemShut {NoStop}%
\bibitem [{\citenamefont {Romero}\ \emph {et~al.}(2024{\natexlab{b}})\citenamefont {Romero}, \citenamefont {Chen}, \citenamefont {Platero},\ and\ \citenamefont {Ban}}]{romero2024optimizing}%
  \BibitemOpen
  \bibfield  {author} {\bibinfo {author} {\bibfnamefont {S.~V.}\ \bibnamefont {Romero}}, \bibinfo {author} {\bibfnamefont {X.}~\bibnamefont {Chen}}, \bibinfo {author} {\bibfnamefont {G.}~\bibnamefont {Platero}},\ and\ \bibinfo {author} {\bibfnamefont {Y.}~\bibnamefont {Ban}},\ }\href {https://doi.org/10.1103/PhysRevApplied.21.034033} {\bibfield  {journal} {\bibinfo  {journal} {Phys. Rev. Appl.}\ }\textbf {\bibinfo {volume} {21}},\ \bibinfo {pages} {034033} (\bibinfo {year} {2024}{\natexlab{b}})}\BibitemShut {NoStop}%
\bibitem [{\citenamefont {Chandarana}\ \emph {et~al.}(2023)\citenamefont {Chandarana}, \citenamefont {Hegade}, \citenamefont {Montalban}, \citenamefont {Solano},\ and\ \citenamefont {Chen}}]{chandarana2023digitized}%
  \BibitemOpen
  \bibfield  {author} {\bibinfo {author} {\bibfnamefont {P.}~\bibnamefont {Chandarana}}, \bibinfo {author} {\bibfnamefont {N.~N.}\ \bibnamefont {Hegade}}, \bibinfo {author} {\bibfnamefont {I.}~\bibnamefont {Montalban}}, \bibinfo {author} {\bibfnamefont {E.}~\bibnamefont {Solano}},\ and\ \bibinfo {author} {\bibfnamefont {X.}~\bibnamefont {Chen}},\ }\href {https://journals.aps.org/prapplied/abstract/10.1103/PhysRevApplied.20.014024} {\bibfield  {journal} {\bibinfo  {journal} {Phys. Rev. Appl.}\ }\textbf {\bibinfo {volume} {20}},\ \bibinfo {pages} {014024} (\bibinfo {year} {2023})}\BibitemShut {NoStop}%
\bibitem [{\citenamefont {Dalal}\ \emph {et~al.}(2024)\citenamefont {Dalal}, \citenamefont {Montalban}, \citenamefont {Hegade}, \citenamefont {Cadavid}, \citenamefont {Solano}, \citenamefont {Awasthi}, \citenamefont {Vodola}, \citenamefont {Jones}, \citenamefont {Weiss},\ and\ \citenamefont {F\"uchsel}}]{dalal2024digitizedcounterdiabaticquantumalgorithms}%
  \BibitemOpen
  \bibfield  {author} {\bibinfo {author} {\bibfnamefont {A.}~\bibnamefont {Dalal}}, \bibinfo {author} {\bibfnamefont {I.}~\bibnamefont {Montalban}}, \bibinfo {author} {\bibfnamefont {N.~N.}\ \bibnamefont {Hegade}}, \bibinfo {author} {\bibfnamefont {A.~G.}\ \bibnamefont {Cadavid}}, \bibinfo {author} {\bibfnamefont {E.}~\bibnamefont {Solano}}, \bibinfo {author} {\bibfnamefont {A.}~\bibnamefont {Awasthi}}, \bibinfo {author} {\bibfnamefont {D.}~\bibnamefont {Vodola}}, \bibinfo {author} {\bibfnamefont {C.}~\bibnamefont {Jones}}, \bibinfo {author} {\bibfnamefont {H.}~\bibnamefont {Weiss}},\ and\ \bibinfo {author} {\bibfnamefont {G.}~\bibnamefont {F\"uchsel}},\ }\href {https://doi.org/10.1103/PhysRevApplied.22.064068} {\bibfield  {journal} {\bibinfo  {journal} {Phys. Rev. Appl.}\ }\textbf {\bibinfo {volume} {22}},\ \bibinfo {pages} {064068} (\bibinfo {year} {2024})}\BibitemShut {NoStop}%
\bibitem [{\citenamefont {Hormozi}\ \emph {et~al.}(2017)\citenamefont {Hormozi}, \citenamefont {Brown}, \citenamefont {Carleo},\ and\ \citenamefont {Troyer}}]{hormozi2017nonstoquastic}%
  \BibitemOpen
  \bibfield  {author} {\bibinfo {author} {\bibfnamefont {L.}~\bibnamefont {Hormozi}}, \bibinfo {author} {\bibfnamefont {E.~W.}\ \bibnamefont {Brown}}, \bibinfo {author} {\bibfnamefont {G.}~\bibnamefont {Carleo}},\ and\ \bibinfo {author} {\bibfnamefont {M.}~\bibnamefont {Troyer}},\ }\href {https://doi.org/10.1103/PhysRevB.95.184416} {\bibfield  {journal} {\bibinfo  {journal} {Phys. Rev. B}\ }\textbf {\bibinfo {volume} {95}},\ \bibinfo {pages} {184416} (\bibinfo {year} {2017})}\BibitemShut {NoStop}%
\bibitem [{\citenamefont {Barkoutsos}\ \emph {et~al.}(2020)\citenamefont {Barkoutsos}, \citenamefont {Nannicini}, \citenamefont {Robert}, \citenamefont {Tavernelli},\ and\ \citenamefont {Woerner}}]{Barkoutsos2020improving}%
  \BibitemOpen
  \bibfield  {author} {\bibinfo {author} {\bibfnamefont {P.~K.}\ \bibnamefont {Barkoutsos}}, \bibinfo {author} {\bibfnamefont {G.}~\bibnamefont {Nannicini}}, \bibinfo {author} {\bibfnamefont {A.}~\bibnamefont {Robert}}, \bibinfo {author} {\bibfnamefont {I.}~\bibnamefont {Tavernelli}},\ and\ \bibinfo {author} {\bibfnamefont {S.}~\bibnamefont {Woerner}},\ }\href {https://doi.org/10.22331/q-2020-04-20-256} {\bibfield  {journal} {\bibinfo  {journal} {{Quantum}}\ }\textbf {\bibinfo {volume} {4}},\ \bibinfo {pages} {256} (\bibinfo {year} {2020})}\BibitemShut {NoStop}%
\bibitem [{\citenamefont {Barron}\ \emph {et~al.}(2024)\citenamefont {Barron}, \citenamefont {Egger}, \citenamefont {Pelofske}, \citenamefont {Bärtschi}, \citenamefont {Eidenbenz}, \citenamefont {Lehmkuehler},\ and\ \citenamefont {Woerner}}]{barron2023provableboundsnoisefreeexpectation}%
  \BibitemOpen
  \bibfield  {author} {\bibinfo {author} {\bibfnamefont {S.~V.}\ \bibnamefont {Barron}}, \bibinfo {author} {\bibfnamefont {D.~J.}\ \bibnamefont {Egger}}, \bibinfo {author} {\bibfnamefont {E.}~\bibnamefont {Pelofske}}, \bibinfo {author} {\bibfnamefont {A.}~\bibnamefont {Bärtschi}}, \bibinfo {author} {\bibfnamefont {S.}~\bibnamefont {Eidenbenz}}, \bibinfo {author} {\bibfnamefont {M.}~\bibnamefont {Lehmkuehler}},\ and\ \bibinfo {author} {\bibfnamefont {S.}~\bibnamefont {Woerner}},\ }\href {https://doi.org/10.1038/s43588-024-00709-1} {\bibfield  {journal} {\bibinfo  {journal} {Nature Computational Science}\ }\textbf {\bibinfo {volume} {4}},\ \bibinfo {pages} {865–875} (\bibinfo {year} {2024})}\BibitemShut {NoStop}%
\bibitem [{\citenamefont {Clausen}(1999)}]{clausen1999branch}%
  \BibitemOpen
  \bibfield  {author} {\bibinfo {author} {\bibfnamefont {J.}~\bibnamefont {Clausen}},\ }\href {https://janders.eecg.toronto.edu/1387/readings/b_and_b.pdf} {\bibfield  {journal} {\bibinfo  {journal} {Department of computer science, University of Copenhagen}\ ,\ \bibinfo {pages} {1}} (\bibinfo {year} {1999})}\BibitemShut {NoStop}%
\bibitem [{\citenamefont {{Qiskit contributors}}(2023)}]{Qiskit}%
  \BibitemOpen
  \bibfield  {author} {\bibinfo {author} {\bibnamefont {{Qiskit contributors}}},\ }\href {https://doi.org/10.5281/zenodo.2573505} {\bibinfo {title} {Qiskit: An open-source framework for quantum computing}} (\bibinfo {year} {2023})\BibitemShut {NoStop}%
\bibitem [{\citenamefont {{IBM Quantum}}(2024)}]{frac}%
  \BibitemOpen
  \bibfield  {author} {\bibinfo {author} {\bibnamefont {{IBM Quantum}}},\ }\href@noop {} {\bibinfo {title} {{New fractional gates reduce circuit depth for utility-scale workloads}}},\ \bibinfo {howpublished} {\url{https://www.ibm.com/quantum/blog/fractional-gates}} (\bibinfo {year} {2024}),\ \bibinfo {note} {[Online: 17/01/25]}\BibitemShut {NoStop}%
\end{thebibliography}%
\clearpage

\end{document}